\documentclass{ifacconf}
\usepackage{graphicx}
\usepackage{natbib}
\usepackage{flushend, cuted}
\usepackage{lipsum}
\usepackage{comment}
\usepackage{colortbl}
\usepackage[centertags]{amsmath}
\usepackage{amsfonts}
\usepackage{newlfont}
\usepackage{mathrsfs}
\usepackage{multirow}
\usepackage{nomencl}
\usepackage{fancyhdr}
\usepackage{arydshln}
\usepackage{tabularx}
\usepackage{booktabs}
\usepackage{wrapfig}
\usepackage{algorithm}
\usepackage{algorithmic}
\usepackage{amssymb}
\usepackage{xcolor}
\begin{document}
\begin{frontmatter}

\title{Eco-driving Trajectory Planning of a Heterogeneous Platoon in Urban Environments\thanksref{footnoteinfo}}

\thanks[footnoteinfo]{This work was partially supported by the US National Science Foundation under award number CNS-1931981.\\© 2022 the authors. This work has been accepted to IFAC for publication under a Creative Commons Licence CC-BY-NC-ND.}

\author[First]{Hao Zhen},
\author[Second]{Sahand Mosharafian},
\author[First]{Jidong J. Yang},
\author[Second]{Javad Mohammadpour Velni}

\address[First]{School of Environmental, Civil, Agricultural \& Mechanical Engineering, University of Georgia, Athens, GA 30602, USA}

\address[Second]{School of Electrical \& Computer Engineering, University of Georgia, Athens, GA 30602, USA}

\begin{abstract}

Given the increasing popularity and demand for connected and autonomous vehicles (CAVs), Eco-driving and platooning in highways and urban areas to increase the efficiency of the traffic system is becoming a possibility. This paper presents Eco-driving trajectory planning for a platoon of heterogeneous electric vehicles (EVs) in urban environments. The proposed control strategy for the platoon considers energy consumption, mobility and passenger comfort, with which vehicles may pass signalized intersections with no stops. For a given urban route, first, the platoon's leader vehicle employs dynamic programming (DP) to plan a trajectory for the anticipated path with the aim of balancing energy consumption, mobility and passenger comfort. Then, every other following CAV in the platoon either follows its preceding vehicle using a PID-based cooperative adaptive cruise control or plans its own trajectory by checking whether it can pass the next intersection without stopping. Furthermore, a heavy-duty vehicle that cannot efficiently follow a light-weight vehicle would instead employ the DP-based trajectory planner. Simulation studies demonstrate the efficacy of the proposed control strategy with which the platoon's energy consumption is shown to reduce while the mobility is not compromised.
\end{abstract}


\end{frontmatter}


\section{Introduction}

\noindent  \textcolor{black}{Driver assistant equipment in modern vehicles has been used to facilitate driving by performing repetitive tasks such as stop and go while reducing fuel consumption and improving safety \citep{moser2015cooperative}. More advanced driver assistance systems are capable of enhancing the traffic system by increasing the road capacity \citep{bengler2014three}. The emerging connected and autonomous vehicle (CAV) technologies offer an unprecedented opportunity to enable Eco-driving and platooning in different traffic environments to maximize the efficiency of traffic systems. It is noted by \cite{huang2018eco} after training drivers for Eco-driving behaviors or using Eco-driving advisory devices in human-driven vehicles, immediate and noticeable reduction on $CO_2$ emission and fuel consumption have been observed while traveling time has slightly increased. However, the Eco-driving skills gained from training are not likely to be sustainable as human driving habits are heterogeneous and formed over years, resulting in diminishing positive impacts on energy consumption over time. Instead, autonomous driving systems can be designed and leveraged to enable enduring and successful Eco-driving.}

\cite{sanguinetti2017many} categorize Eco-driving behaviors into six classes: route planning, driving behavior, maintenance, comfort, fueling and load management. Specifically, driving behavior can be further divided into different categories, namely accelerating and decelerating, cruising, waiting, parking and driving mode selection \citep{huang2018eco}. Our study leverages autonomous driving to improve energy consumption and mobility by focusing on two specific aspects: route planning and driving behavior. It was noted that although Eco-driving might not negatively impact safety, it could result in driving with relatively low speed, and thereby increase traveling time \citep{wu2015energy}. As such, recent studies have been focused on balancing energy consumption and mobility \citep{oh2018eco,ma2021eco}. Different from the existing research being concentrated on Eco-driving in different mixed traffic conditions over a short distance (e.g., see \citep{zhao2018platoon}), the focus of our study is on long-term trajectory planning for a platoon of heterogeneous electric vehicles (EVs) considering energy consumption, mobility and passenger comfort in urban environments.

Connected and autonomous vehicle (CAV) technologies are not only able to enhance sustainability but are also capable of improving safety and mobility \citep{wang2019cooperative}. Wireless communication among CAVs enables cooperative adaptive cruise control (CACC), which allows safe driving with a smaller gap between vehicles \citep{mosharafian2021gaussian, wang2015string,kreuzen2012cooperative}. Different control methods are used to implement CACC, namely model predictive control (MPC)-based methods and proportional–derivative /proportional–integral–derivative (PD/PID) controllers. It has been shown that Eco-driving for vehicles employing the knowledge about traffic signal timing can further decrease energy consumption in the urban environment compared to the situation without such information \citep{teichert2020comparison}, in which unnecessary braking and idling at intersections can increase energy consumption \citep{oh2018eco}. In the existing literature, the impact of combining Eco-driving and CACC at signalized intersections (a.k.a. Eco-CACC) has been investigated \citep{yang2016eco,wang2019cooperative}. While some previous works considered single-intersection scenarios, others studied a corridor setting that includes multiple intersections. Although \cite{wang2019cooperative} studied a scenario including numerous signalized intersections, their Eco-driving planning is only conducted for the upcoming intersection but not the entire path. \cite{ma2021eco} considered Eco-driving-based CACC in a path with multiple intersections while planning the speed trajectory for the entire route. However, their proposed solution is not useful for those vehicles in the platoon that cannot pass the traffic light. In this situation, the leader would slightly change its trajectory to allow all following vehicles to pass the traffic light during the green phase. This would result in a sub-optimal solution, even for the vehicles that can pass the traffic light following the claimed "optimal" trajectory. Moreover, this approach would no longer work for a longer platoon, as stopping at the traffic light for some vehicles is inevitable.

Heterogeneity of the platoon may affect the platoon's stability, efficiency and mobility. Some researchers have been studying the heterogeneity from different aspects, e.g., heterogeneity of mixed platoon has been studied by modeling the interaction between human-driven vehicles and CAVs \citep{jia2018multiclass}. Some authors have considered the Eco-CACC of a heterogeneous platoon of heavy-duty vehicles with a time delay from the perspective of internal stability and string stability \citep{zhai2020ecological}. However, there is less concentration on the trajectory planning of heterogeneous platoons, to the best of our knowledge. In this paper, the Eco-CACC problem is studied for a platoon of heterogeneous electric vehicles in an urban environment setting. The traveling route for the platoon includes multiple signalized intersections whose signal timings are available to the platoon through Vehicle-to-infrastructure (V2I) communication. The proposed trajectory planning for vehicles allows each CAV to either follow its predecessor using a PID-based CACC approach or use the route, signal and traffic information to re-design its own trajectory instead, which is found by solving a dynamic programming (DP) problem. The proposed approach allows all CAVs to pass traffic lights with no stops while balancing energy consumption, mobility and passenger comfort.

The contributions of the paper are summarized as follows.
\begin{enumerate}
    \item An Eco-CACC strategy is proposed for a platoon of heterogeneous electric vehicles in urban environments, a general scheme that can be adapted to various situations such as different platoon sizes and different route settings.
    \item Heterogeneity of a platoon is considered by specifying light-weight vehicles versus heavy-duty vehicles, which exhibit different performance characteristics.
    \item The global optimal trajectory of leading vehicles is determined by solving an optimization problem using DP algorithm aiming at striking a balance between energy consumption, mobility and comfort while other vehicles employ PID to follow their preceding vehicle.
\end{enumerate}

The remainder of the paper is organized as follows. The vehicle model considered in this study is explained in Section 2. The proposed Eco-CACC strategy is discussed in Section 3. Simulation results and discussions are presented in Section 4, and finally, concluding remarks are provided in Section 5.


\section{System Model}

\begin{figure*}[htp!]
  \centering
    \includegraphics[width=0.9\textwidth]{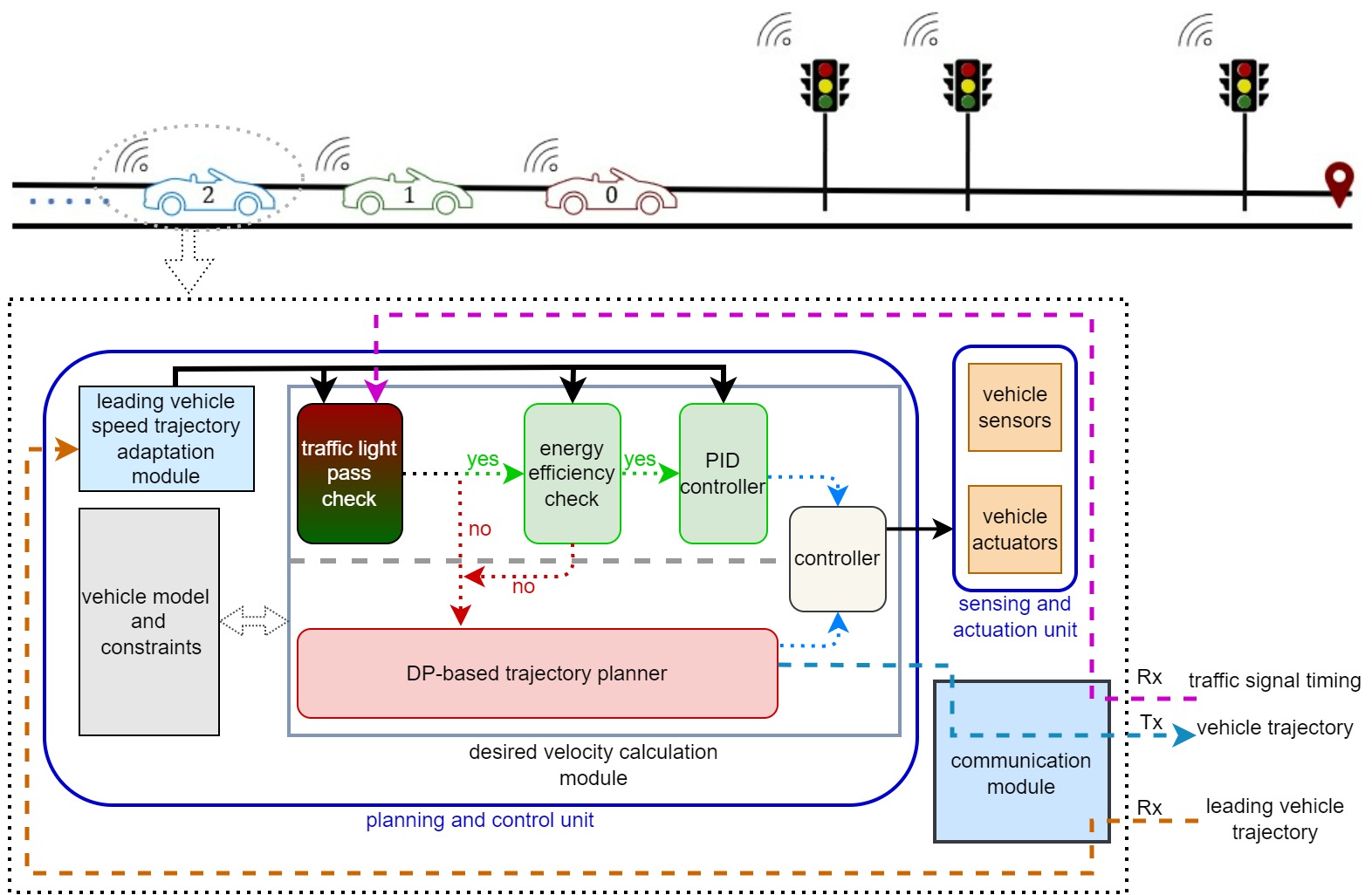}
  \caption{Diagram of the proposed Eco-CACC design. In this structure, a platoon of multiple CAVs is considered, traveling along an arterial with
  several signalized intersections. Each vehicle plans the trajectory itself or follows its preceding vehicle
  depending on the circumstances (whether the CAV can pass the intersections following its preceding vehicle or if its average energy efficiency through the whole route is lower than a predefined threshold). For example, a heavy-duty vehicle that cannot efficiently follow its predecessor will use the DP-based trajectory planner instead.}
  \label{fig:diagram}
\end{figure*}

\noindent In this paper, we address the problem of finding an Eco-CACC strategy for a heterogeneous CAV platoon traveling along an urban arterial with multiple traffic lights. The following assumptions are made throughout the paper:
\begin{enumerate}
    \item Only longitudinal control of CAVs is considered along an arterial.
    \item There is no overtaking, and other vehicles do not cut in the CAV platoon.
    \item All CAVs in platoon are also EVs and have access to and will leverage the traffic signal phase and timing (SPaT) information through V2I communication.
\end{enumerate}
The diagram of the proposed Eco-CACC scheme is presented in Fig. \ref{fig:diagram}. The remainder of this section describes the vehicle dynamics and energy consumption model used in our study.

\color{black}

\subsection{Vehicle Dynamics}

In this paper, a simplified longitudinal vehicle dynamics model is used as
\begin{equation}\label{v_dynamics}
    \begin{gathered}
    m\,\frac{dv}{dt}=F_t-f,
    \end{gathered}
\end{equation}
where $m$, $v$, $F_t$ and $f$ are the vehicle's total mass, velocity, total traction force and total resistance, respectively. The vehicle's total resistance is calculated as follows
\begin{multline}\label{total_resistance}
f = F_{a}+F_{r l}+F_{g} \\ = 0.5 C_{d} A \rho v^{2}+\mu m g \cos \theta+m g \sin \theta,
\end{multline}
where $F_a$, $F_{rl}$ and $F_g$ are aerodynamic resistance, rolling resistance and grade resistance, respectively. The constant $C_{d}$ represents drag coefficient, $A$ is the projected frontal area of the vehicle, $\rho$ is the air density, $\mu$ is the friction coefficient and $\theta$ is the road grade, which is set to zero in this paper.

The required torque $T$ can be calculated using \eqref{v_dynamics} and \eqref{total_resistance}, thereby
\begin{equation}\label{torque}
    \begin{gathered}
    T=\frac{r}{\epsilon_0\,\eta_d}(0.5\,C_d\,A\,\rho\,v^2+\mu\,m\,g\,\cos{\theta}+m\,g\,\sin{\theta}+m\,a),
    \end{gathered}
\end{equation}
where $r$ is the wheel's radius, $\epsilon_0$ is the gear reduction ratio and $\eta_d$ is the mechanical efficiency of the drivetrain.

\subsection{Electric Vehicle Energy consumption Model}

The following comprehensive power-based electric vehicle's energy consumption model is used to calculate energy consumption \citep{fiori2016power}: 
\begin{equation}
    \begin{gathered}
    P=(m\,a+\mu\,m\,g\,\cos{\theta}+0.5\,C_D\,\rho\,A\,v^2+m\,g\,\sin{\theta})\,v.
    \end{gathered}
\end{equation}
Thus, the energy consumed by an electric vehicle, denoted by $E$, over a time span of $\Delta t$ is calculated as
\begin{equation}
    \begin{gathered}
    E=
    \begin{cases}
    (P\Delta t/3600)/\eta_{tr}\, & \text{Traction};\\
    (P\Delta t/3600)/\eta_{re}\, & \text{Recuperation},
    \end{cases}
    \end{gathered}
\end{equation}
where $\eta_{tr}$ is the motor efficiency and  $\eta_{re}$ is the recuperation efficiency. In this paper, the motor efficiency $\eta_{tr}$ for every step is interpolated from the motor map, which is a lookup table containing the motor efficiency values for different combinations of torque and rotation speed.


\section{Eco-CACC Strategy}

In this section, we provide details on our proposed approach for Eco-driving trajectory planning, as well as a PID-based CACC scheme.

\subsection{DP-based Eco-driving Trajectory Planning}

The proposed trajectory planning approach is formulated as a constrained optimization problem and solved using dynamic programming (DP). The objective is to find the optimal trajectory profile for the leader vehicle along an urban route to balance the energy consumption, mobility and passenger comfort and enable the vehicle to pass intersections without stopping. It should be noted that the proposed algorithm is distance-based, and so the entire route is discretized into $N$ distance steps. Variable $s_k$ denotes the $k^{\text{th}}$ step distance. The state variables include the velocity and time, and the control variable is acceleration that governs state space changes. In addition, the following constraints need to be considered based on the vehicle's acceleration limitation, bound on traveling time, and the road speed limit:
\begin{equation*}
    \begin{gathered}
    v_k\in[0,v_{\max,k}],\\
t_k\in[t_{\min,k},t_{\max,k}],\\
a_k\in[a_{min},a_{max}],
    \end{gathered}
\end{equation*}
where variable $v_k$ denotes the velocity in $k^{\text{th}}$ step, $t_k$ denotes the time span in $k^{\text{th}}$ step, and $a_k$ denotes the acceleration in $k^{\text{th}}$ step. Hence, the optimization problem is formulated as
\begin{equation}
\begin{gathered}\label{DP}
    \min_{a_k} \sum_{k=1}^{N}[\alpha E(v_k,a_k)+\beta M(v_k)+\gamma C(a_k)]+\mu P_{red},\\
    \text{subject to:\hspace{180pt}}\\
    M(v_k)=(ds/(v_k+0.01)-ds/v_{des})^2,\\
    C(a_k)=a_k^2,\\
    \mu =
    \begin{cases} 1\,\,;\quad t_{k}|_{v_{k}, a_{k}, k = \delta} \in[t_{red}, t_{green}),\\
    0\,\,;\quad t_{k}|_{v_{k}, a_{k}, k = \delta} \in[t_{green}, t_{red})\; or \; k\neq \delta,
    \end{cases}\\
v_{k+1}=\sqrt{v_k^2+2\,a_k\,(s_{k+1}-s_k)},\\
t_{k+1}=t_k+2\,(s_{k+1}-s_k)/(v_{k+1}+v_k),
\end{gathered}
\end{equation}
where $E(v_k,a_k)$ denotes $k^{\text{th}}$ step energy consumption given $v_k$ and $a_k$. $M(v_k)$ represents the mobility, in which $v_{des}$ denotes the desired velocity along the urban route and parameter $ds$ represents the distance step and is set to $1\,m$ in this paper.
$C(a_k)$ denotes the passenger comfort defined as the square of acceleration \citep{oh2018eco}. $\alpha$, $\beta$ and $\gamma$ are the weights for energy consumption, mobility and passenger comfort, respectively. Furthermore, $P_{red}$ is the penalty for the case that a vehicle runs into red light at the traffic signal. Parameter $\mu$ determines whether or not to add the penalty $P_{red}$. $t_{green}$ denotes the moment at which the traffic light switches from red to green while $t_{red}$ denotes the moment when the traffic light switches from green to red. According to aforementioned definitions, the interval $[t_{green},t_{red})$ denotes the traffic light's green phase (effective green) while $[t_{red},t_{green})$ denotes the traffic light's red phase. Note the green phase and red phase refer to effective green duration and effective red duration that absorb the yellow interval.  Furthermore, $\delta$ is used to denote the number of distance steps to arrive at traffic light from the current vehicle position. 

\subsection{PID-based CACC Scheme}

Due to the computational complexity of the DP algorithm, it is used to find the optimal Eco-driving trajectory profile of the leader vehicle(s) while the following vehicles in the platoon aim to follow their respective preceding vehicles at the desired distance $d_{r,i}$ with the same speed as their predecessor using a PID controller. Let the platoon includes $m$ vehicles, with $d_{i}$ denoting the inter-distance between a vehicle $i$ and its preceding vehicle $i-1$ and $v_{i}$ the velocity of vehicle $i$. A constant-time spacing policy is adopted for the following vehicle as follows
\begin{equation}
d_{r, i}(t)=r_{i}+h v_{i}(t), \quad 2<i<m
\end{equation}
where $d_{r,i}(t)$ is the desired inter-distance of vehicle $i$ from vehicle $i-1$, $h$ is the time headway constant, and $r_{i}$ is the standstill distance. The following vehicle model is considered
\begin{equation}
\left(\begin{array}{c}
\dot{d}_{i} \\
\dot{v}_{i} \\
\dot{a}_{i}
\end{array}\right)=\left(\begin{array}{c}
v_{i-1}-v_{i} \\
a_{i} \\
-\frac{1}{\tau} a_{i}+\frac{1}{\tau} u_{i}
\end{array}\right), \quad 2 \leq i \leq m,
\end{equation}
where $a_{i}$ is the $i^{\text{th}}$ vehicle’s acceleration, $u_{i}$ is the input, and $\tau$ is the time constant of driveline dynamics. The spacing error $e_{i}(t)$, which is the difference between the current inter-vehicle distance and its desired value, is calculated as
\begin{equation}
\begin{aligned}
e_{i}(t) &=d_{i}(t)-d_{r, i}(t) \\
&=\left(s_{i-1}(t)-s_{i}(t)-L_{i}\right)-\left(r_{i}+h v_{i}(t)\right),
\end{aligned}
\end{equation}
where $s_{i}(t)$ is the $i^{\text{th}}$ vehicle’s position and $L_{i}$ is its length. The control action for the following vehicle is given as \citep{ploeg2011design}
\begin{equation}
\dot{u}_{i}=-\frac{1}{h} u_{i}+\frac{1}{h}\left(k_{p} e_{i}+k_{d} \dot{e}_{i}+k_{a} \ddot{e}_{i}\right)+\frac{1}{h} u_{i-1},
\end{equation}
where $K=\left[\begin{array}{lll}
k_{p} & k_{d} & k_{a}
\end{array}\right]$ are the gains of the PID feedback controller. In our simulation studies (see Section 4), the PID gains are set to be $K=[0.001~~ 10~~ 1]$ for both scenarios examined.

\subsection{Proposed Eco-CACC Strategy}

Our ultimate goal of designing the control strategy for a platoon is to find an optimal solution for the entire platoon, which allows all vehicles in the original platoon to pass all traffic lights with no stops while maintaining a balance between the platoon’s energy consumption, mobility and passenger comfort. However, the optimal trajectory for the leading vehicle in a heterogeneous platoon is not necessarily optimal for the remainder of the CAVs due to the traffic light phase changes or vehicle heterogeneity. Instead, a control strategy is introduced for each vehicle in the platoon to reach a near-optimal solution. Each follower vehicle may decide to either follow its predecessor using a PID-based CACC scheme or employ DP to find the proper trajectory for itself.

\begin{figure}[t!]
\begin{center}
\includegraphics[width=\linewidth]{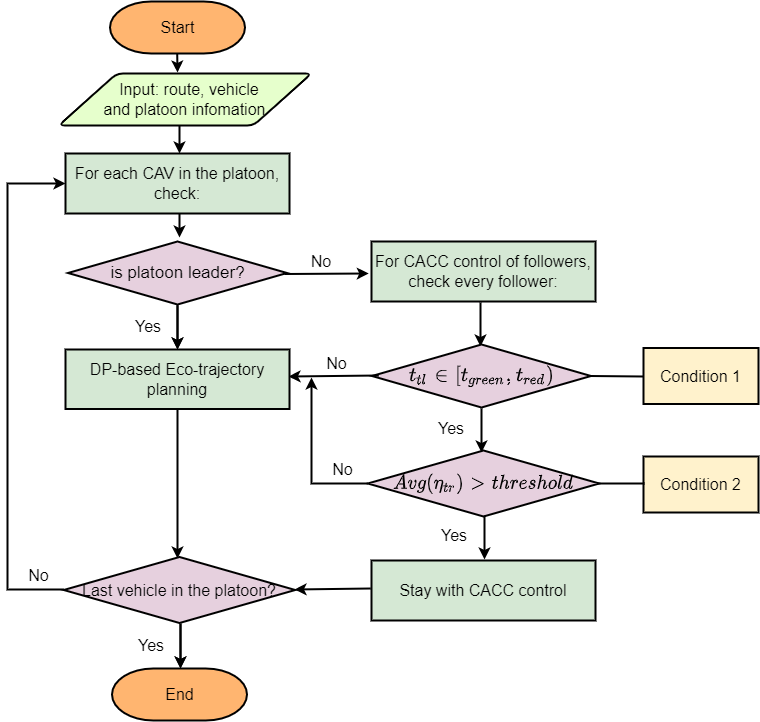}
\caption{The proposed control strategy for a heterogeneous platoon in urban environments.}
\label{fig:control strategy}
\end{center}
\end{figure}

The block diagram showing details of the proposed strategy is presented in Fig. \ref{fig:control strategy}. The control strategy uses route information, including traffic light location and SPaT information. The trajectory planning for the leader vehicle is implemented using DP to balance energy consumption, mobility and passenger comfort. After the leader's trajectory planning, PID-based CACC is employed for each follower vehicle in light of the low computational cost. However, two conditions need to be considered for a long heterogeneous platoon. First, if the onset of red light stops one or more follower vehicles, the energy consumption will increase due to the deceleration and stopping of those vehicles. Thus, each follower CAV checks 'Condition 1', that is, whether or not it can pass all the intersections at green when following its preceding vehicle; if 'Condition 1' is not satisfied, then the CAV will be set as a 'New Leader', and its trajectory will be re-planned using DP. Second, because of the heterogeneity of CAVs in the platoon, the leader vehicle's optimal trajectory is not necessarily suitable for all the follower vehicles. Thus, each follower vehicle in the platoon checks whether or not it satisfies this 'Condition 2'; if the average efficiency $Avg(\eta_{tr})$ falls below a preset threshold, then it becomes a 'New Leader' and will plan its trajectory using DP. This Eco-CACC control strategy applies to all vehicles in the platoon.

\section{Simulation Results and Discussion}

\noindent A platoon containing 20 heterogeneous electric vehicles is considered in the simulation study. Among the $20$ vehicles, the $16^{\text{th}}$ vehicle is designated as a heavy-duty vehicle denoted by HC. The parameters of all CAVs are presented in Table \ref{table:1}, where
'Light' refers to light-weight vehicles, and 'Heavy' refers to heavy-duty vehicles. Besides the mass, the frontal area of the vehicle, aerodynamic drag coefficient \textcolor{black}{and motor efficiency map}, the difference between the light-weight vehicles and the heavy-duty vehicle also includes the maximum acceleration, which is set to $4\,m^{2} /s$ for the light-weight vehicles and $2.5\,m^{2} /s$ for the heavy-duty vehicle.

A 2500-meter-long urban arterial with two pre-timed traffic lights located at distances $600\,m$ and $2000\,m$ along the route is considered to evaluate the performance of the proposed Eco-CACC strategy. The first traffic light at $600\,m$ 
has 72-second effective green for the through phase along the route, followed by 88-second effective red. The second traffic light at $2000\,m$ has 
75-second effective green, followed by 95-second effective red. The maximum speed of the arterial 
is $60\,km/h$. The distance step $ds$ for this simulation is set to $1\,m$. 

\begin{table}[t]
\centering\tiny
\caption{Electric vehicle model parameters used in the simulation study.}
\begin{tabular}{|p{0.97cm}|p{1.03cm}|p{1.03cm}|p{.97cm}|p{1.05cm}|p{1.05cm}|} 
 \hline
Parameter & Light & Heavy & Parameter & Light & Heavy\\ [0.5ex]
 \hline\hline
 $m$ & $1400kg$ & $1900kg$ & $C_d$ & $0.36$&$0.7$ \\ \hline
 $\mu$ & $0.008$& $0.008$ & $\rho$ & $1.2\,kg\,m^{-3}$&$1.2\,kg\,m^{-3}$\\ \hline
 $A$ & $4.5\,m^2$ & $8.5\,m^2$& $\epsilon_0$ & $3.92$ & $3.92$\\ \hline
 $r$ & $0.282\,m$ & $0.282\,m$ & $\eta_d$ & $0.95$& $0.95$\\ [1ex] 
 \hline
\end{tabular}
\label{table:1}
\end{table}

Two scenarios are evaluated. In Scenario \uppercase\expandafter{\romannumeral1}, only vehicle mobility and passenger comfort are considered while Scenario \uppercase\expandafter{\romannumeral2} considers energy consumption as well.

\subsection{Eco-CACC Considering Mobility and Passenger Comfort (Scenario \uppercase\expandafter{\romannumeral1})}

\begin{figure*}[htp!]
\begin{center}
\includegraphics[width=\linewidth]{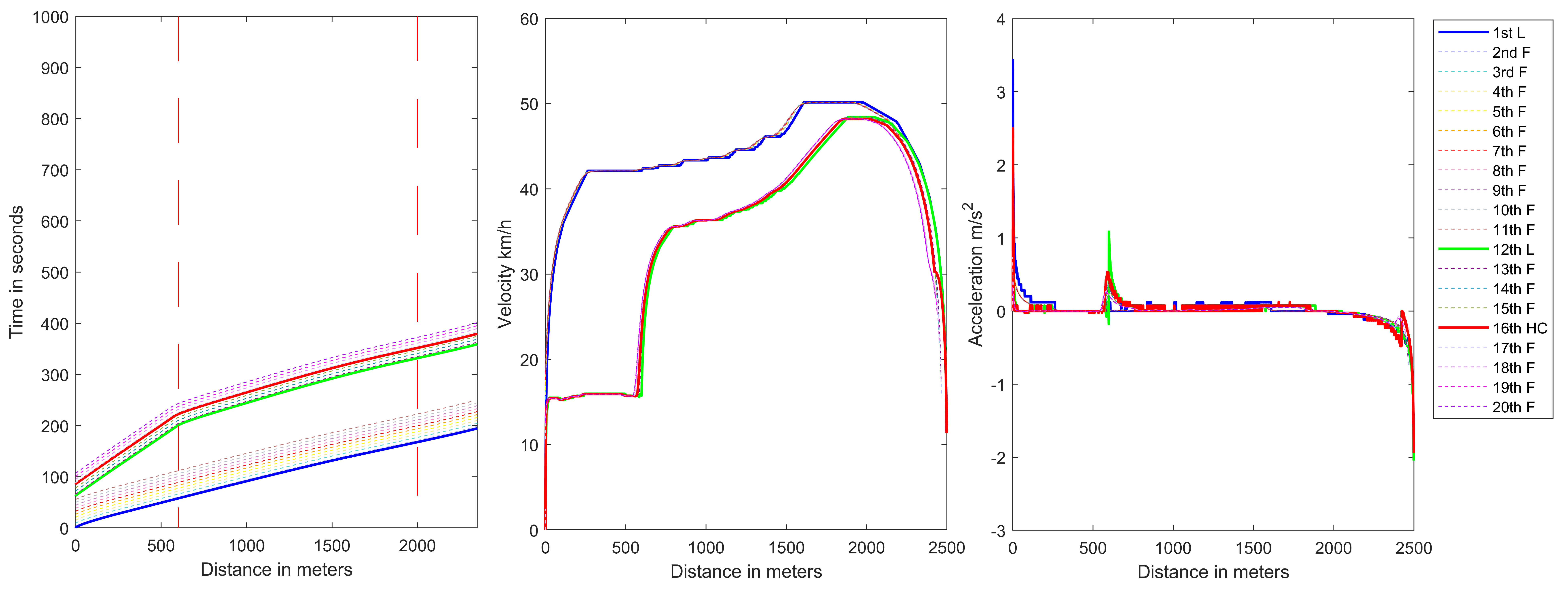}
\caption{Eco-driving trajectory planning of heterogeneous platoon in Scenario \uppercase\expandafter{\romannumeral1}. The average traveling time for this scenario is $234.86\,s/veh$ and the maximum velocity in the platoon is below $51\,km/h$.}
\label{fig:result_M}
\end{center}

\end{figure*}

In this scenario, only vehicle mobility and passenger comfort are considered when carrying out the DP-based trajectory planning for leader vehicles (i.e., $\alpha=0$ in $(6)$). By implementing the proposed Eco-CACC strategy for the entire heterogeneous platoon, location, velocity and acceleration profiles for each vehicle are shown in the first, second and third subplots in Fig. \ref{fig:result_M}, respectively. Letter 'L' denotes leader(s) in the platoon whose trajectories are planned using DP, and letter 'F' indicates follower(s) in the platoon, whose trajectories are obtained through the PID-based CACC scheme. Using the proposed Eco-CACC strategy, the trajectory of the leader vehicle '$1st\,L$' is determined using the DP algorithm; then vehicles '$2nd\,F$' - '$11th\,F$' follow their respective predecessors' trajectories using the PID-based CACC controller. Checking Condition 1 for the $12th$ vehicle indicates that it cannot pass at green if it follows its predecessor using the PID controller. Thus, $12^{\text{th}}$ CAV becomes a new leader and is then renamed as '$12th\,L$', and $13^{\text{th}}$-$15^{\text{th}}$ follower vehicles ('$13th\,F$'-'$15th\,F$') follow it. By checking Condition 2, the trajectory for $16^{\text{th}}$ vehicle ('$16th\,HC$'), which is a heavy-duty vehicle, is re-planned, and vehicles '$17th\,F$'-'$20th\,F$' follow $16th\,HC$'s trajectory by employing a PID controller.

\begin{figure*}[htp!]
\begin{center}
\includegraphics[width=\linewidth]{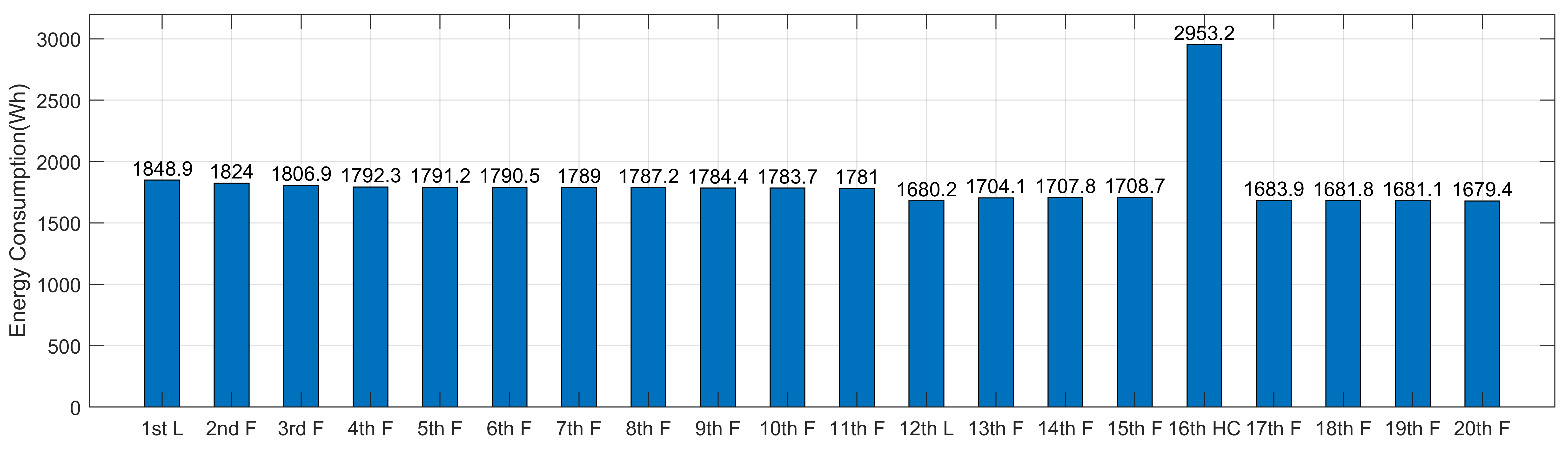}
\caption{Energy consumption for each vehicle through the urban route in Scenario \uppercase\expandafter{\romannumeral1}. Energy consumption for the heavy-duty vehicle is relatively high when the cost in \eqref{DP} does not include the energy consumption term. Hence, the energy consumption should be taken into account while solving \eqref{DP} to improve the overall energy efficiency.}
\label{fig:M_consumption}
\end{center}
\end{figure*}

When only mobility and passenger comfort are considered in the objective function (i.e., in \eqref{DP}), the average traveling time is $234.86\,s/veh$, while the maximum allowed traveling time is set to $1000\,s$. The maximum velocity in the platoon reaches $51\,km/h$. We note that the accelerations for the light-weight vehicles remain below the maximum acceleration limit of $3.5\,m/s^2$ while the heavy-duty vehicle reaches the maximum acceleration limit of $2.5\,m/s^2$. 
The energy consumption of each vehicle in Scenario \uppercase\expandafter{\romannumeral1} is shown in Fig. \ref{fig:M_consumption}. In this scenario, although the traveling time for vehicles is relatively short, the energy aspect was completely disregarded, 
leading to relatively high energy consumption. Besides, the energy consumption for $16th\,HC$ is much higher ($2,953.2\,Wh$) than other vehicles in the platoon that all consumed less than $1850\,Wh$ of energy along the entire route. To balance mobility and sustainability, Scenario \uppercase\expandafter{\romannumeral2} is then studied.

\subsection{Eco-CACC Considering Energy Consumption, Mobility and Passenger Comfort (Scenario \uppercase\expandafter{\romannumeral2})}

\begin{figure*}[htp!]
\begin{center}
\includegraphics[width=\linewidth]{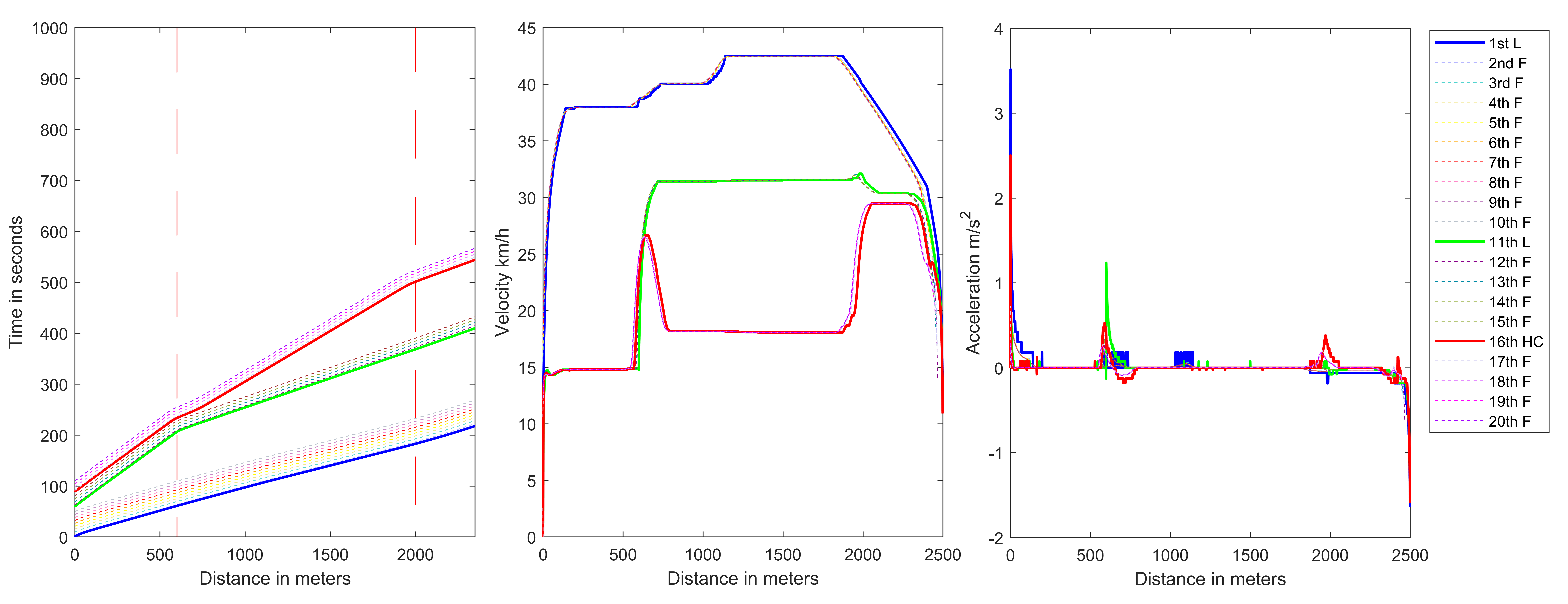}
\caption{Eco-driving trajectory planning of heterogeneous platoon in scenario \uppercase\expandafter{\romannumeral2}. The average traveling time for this scenario is $305.85\,s/veh$ and the maximum velocity in the platoon is about $42.5\,km/h$. In this scenario, the platoon's mobility and energy efficiency are balanced.}
\label{fig:result_EMH}
\end{center}
\end{figure*}

In this scenario, besides the mobility and driver comfort aspects, energy consumption is also considered when carrying out the Eco-trajectory planning. Fig. \ref{fig:result_EMH} depicts the results of the second scenario. Using the proposed Eco-CACC strategy, the trajectory of the leading vehicle '$1st\,L$' is optimized using DP, and the follower vehicles '$2nd\,F$'-'$10th F$' follow their respective predecessor's trajectory using a PID-based CACC scheme. Then, checking Condition 1 for the $11^{\text{th}}$ vehicle indicates that it cannot pass at green if it follows its predecessor using the PID controller. Thus, $11^{\text{th}}$ vehicle becomes a new leader and is hence renamed as '$11th\,L$', and $12^{\text{th}}$- $15^{\text{th}}$ follower vehicles ('$12th\,F$'-'$15th\,F$') follow it afterward. After checking Condition 2 for $16^{\text{th}}$ vehicle (HC), which is a heavy-duty vehicle, its trajectory is re-planned, and its follower vehicles '$17th\,F$'-'$20th\,F$' then follow its trajectory afterward using the PID-based CACC scheme.

\begin{figure*}[htp!]
\begin{center}
\includegraphics[width=18cm]{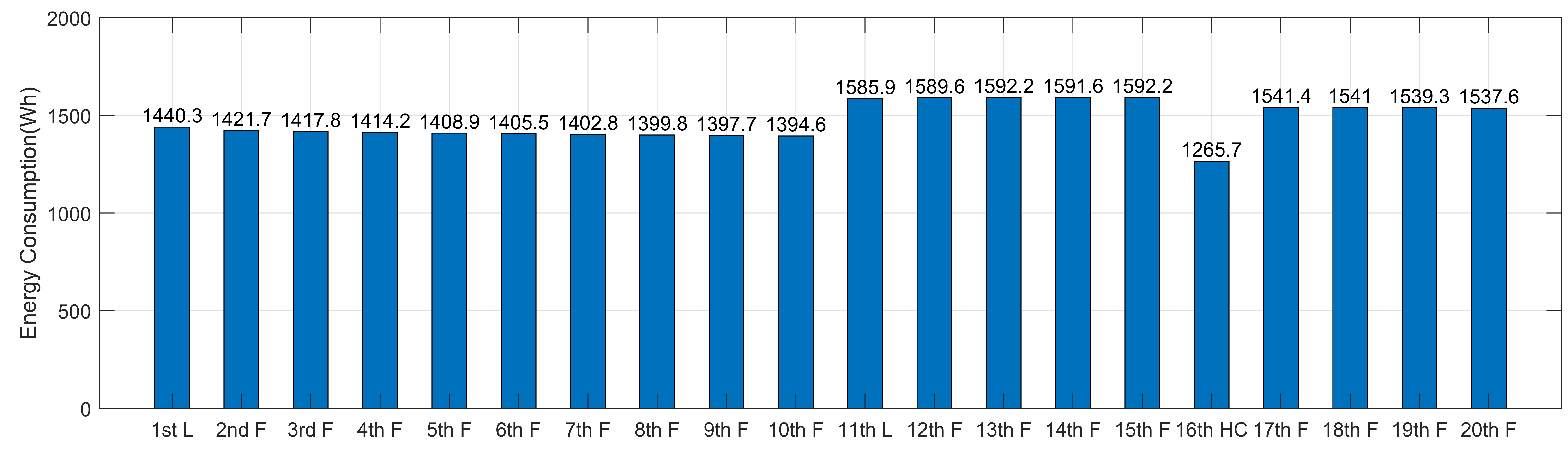}
\caption{Energy consumption for each vehicle through the urban route in scenario \uppercase\expandafter{\romannumeral2}. Energy consumption for CAVs is relatively higher compared to the first scenario due to the mobility term added in \eqref{DP}.}
\label{fig:EMH_consumption}
\end{center}
\end{figure*}

With energy consumption considered in this scenario, the average total traveling time is $305.85\,s/veh$, while the maximum allowed traveling time is set to $1000\,s$. The maximum velocity reaches $42\,km/h$ and the maximum acceleration reaches $3.5\,m/s^2$. In order to maintain a desired velocity to balance its energy consumption, mobility and passenger comfort while passing the green phase, heavy-duty vehicle '$16th HC$' exhibits oscillatory (acceleration and deceleration) behavior around both intersections. Noticeable energy consumption reduction for the heavy-duty vehicle $16th\,HC$ is observed from $2953.2\,Wh$ in Scenario \uppercase\expandafter{\romannumeral1} to $1265.7\,Wh$ in Scenario \uppercase\expandafter{\romannumeral2}. Each vehicle's energy consumption in this scenario is shown in Fig. \ref{fig:EMH_consumption}. The comparison of average energy consumption and average traveling time between Scenario \uppercase\expandafter{\romannumeral1} and Scenario \uppercase\expandafter{\romannumeral2} is shown in Table \ref{table:2comparison}. Compared with Scenario \uppercase\expandafter{\romannumeral1}, the platoon in Scenario \uppercase\expandafter{\romannumeral2} sacrifices about $71\,s$ on average (about 30\%) to trade with about $339\,Wh/veh$ energy consumption reduction (about 19\%). It is noted that the traveling time is not only related to vehicle characteristics but also traffic signal timings and travel start time of vehicles since part of the cost function considers whether a vehicle can pass at the green phase of the traffic light.

\begin{table}
\caption{Comparison between Scenario \uppercase\expandafter{\romannumeral1} and Scenario \uppercase\expandafter{\romannumeral2}}
\begin{tabular}{|c|c|c|}
\hline
\multicolumn{1}{|l|}{Scenario} & \multicolumn{1}{l|}{average energy consumption} & \multicolumn{1}{l|}{average traveling time} \\ \hline
\uppercase\expandafter{\romannumeral1}                              & 1812.965 Wh/veh                                   & 234.86 s/veh                          \\ \hline
\uppercase\expandafter{\romannumeral2}                             & 1473.99 Wh/veh                                  & 305.85 s/veh                         \\ \hline
\end{tabular}
\label{table:2comparison}
\end{table}

\section{Conclusion}
\vspace{-4pt}
\noindent This paper proposes an Eco-CACC strategy for a platoon of heterogeneous electric vehicles in urban environments. The proposed method combines Eco-trajectory planning and CACC, resulting in a near-optimal solution for long platoons along an arterial by balancing energy consumption, mobility and passenger comfort. The trajectory planning problem for the leader vehicle(s) is solved using dynamic programming while follower vehicles employ a PID controller to follow their respective predecessors. The proposed approach also considers traffic SPaT information while solving the dynamic programming problem. With the proposed control strategy, vehicles avoid stopping at traffic lights, thereby improving the energy economy of the entire platoon. Our simulation study shows that adding energy consumption component in the objective function helps to balance mobility and energy consumption of the platoon, resulting in lower energy consumption with slightly increased traveling time. It is noted that due to our discrete distance space when formulating the problem, the acceleration profile may exhibit an oscillatory behavior without the passenger comfort term in the objective function.

\bibliography{ifacconf}

\begin{thebibliography}{18}
\providecommand{\natexlab}[1]{#1}
\providecommand{\url}[1]{\texttt{#1}}
\providecommand{\urlprefix}{URL }
\expandafter\ifx\csname urlstyle\endcsname\relax
  \providecommand{\doi}[1]{doi:\discretionary{}{}{}#1}\else
  \providecommand{\doi}{doi:\discretionary{}{}{}\begingroup
  \urlstyle{rm}\Url}\fi

\bibitem[{Bengler et~al.(2014)Bengler, Dietmayer, Farber, Maurer, Stiller, and
  Winner}]{bengler2014three}
Bengler, K., Dietmayer, K., Farber, B., Maurer, M., Stiller, C., and Winner, H.
  (2014).
\newblock Three decades of driver assistance systems: Review and future
  perspectives.
\newblock \emph{IEEE Intelligent transportation systems magazine}, 6(4), 6--22.

\bibitem[{Fiori et~al.(2016)Fiori, Ahn, and Rakha}]{fiori2016power}
Fiori, C., Ahn, K., and Rakha, H.A. (2016).
\newblock Power-based electric vehicle energy consumption model: Model
  development and validation.
\newblock \emph{Applied Energy}, 168, 257--268.

\bibitem[{Huang et~al.(2018)Huang, Ng, Zhou, Surawski, Chan, and
  Hong}]{huang2018eco}
Huang, Y., Ng, E.C., Zhou, J.L., Surawski, N.C., Chan, E.F., and Hong, G.
  (2018).
\newblock Eco-driving technology for sustainable road transport: A review.
\newblock \emph{Renewable and Sustainable Energy Reviews}, 93, 596--609.

\bibitem[{Jia et~al.(2018)Jia, Ngoduy, and Vu}]{jia2018multiclass}
Jia, D., Ngoduy, D., and Vu, H.L. (2018).
\newblock A multiclass microscopic model for heterogeneous platoon with
  vehicle-to-vehicle communication.
\newblock \emph{Transportmetrica B: Transport Dynamics}.

\bibitem[{Kreuzen(2012)}]{kreuzen2012cooperative}
Kreuzen, C. (2012).
\newblock Cooperative adaptive cruise control: using information from multiple
  predecessors in combination with {MPC}.
\newblock \emph{Delft University of Technology}.

\bibitem[{Ma et~al.(2021)Ma, Yang, Wang, Li, Wu, Zhao, Wu, Aksun-Guvenc, and
  Guvenc}]{ma2021eco}
Ma, F., Yang, Y., Wang, J., Li, X., Wu, G., Zhao, Y., Wu, L., Aksun-Guvenc, B.,
  and Guvenc, L. (2021).
\newblock Eco-driving-based cooperative adaptive cruise control of connected
  vehicles platoon at signalized intersections.
\newblock \emph{Transportation Research Part D: Transport and Environment}, 92,
  102746.

\bibitem[{Moser et~al.(2015)Moser, Waschl, Kirchsteiger, Schmied, and
  Del~Re}]{moser2015cooperative}
Moser, D., Waschl, H., Kirchsteiger, H., Schmied, R., and Del~Re, L. (2015).
\newblock Cooperative adaptive cruise control applying stochastic linear model
  predictive control strategies.
\newblock In \emph{2015 European Control Conference (ECC)}, 3383--3388. IEEE.

\bibitem[{Mosharafian et~al.(2021)Mosharafian, Razzaghpour, Fallah, and
  Velni}]{mosharafian2021gaussian}
Mosharafian, S., Razzaghpour, M., Fallah, Y.P., and Velni, J.M. (2021).
\newblock Gaussian process based stochastic model predictive control for
  cooperative adaptive cruise control.
\newblock In \emph{2021 IEEE Vehicular Networking Conference (VNC)}, 17--23.
  IEEE.

\bibitem[{Oh and Peng(2018)}]{oh2018eco}
Oh, G. and Peng, H. (2018).
\newblock Eco-driving at signalized intersections: What is possible in the
  real-world?
\newblock In \emph{2018 21st International Conference on Intelligent
  Transportation Systems (ITSC)}, 3674--3679. IEEE.

\bibitem[{Ploeg et~al.(2011)Ploeg, Scheepers, Van~Nunen, Van~de Wouw, and
  Nijmeijer}]{ploeg2011design}
Ploeg, J., Scheepers, B.T., Van~Nunen, E., Van~de Wouw, N., and Nijmeijer, H.
  (2011).
\newblock Design and experimental evaluation of cooperative adaptive cruise
  control.
\newblock In \emph{2011 14th International IEEE Conference on Intelligent
  Transportation Systems (ITSC)}, 260--265. IEEE.

\bibitem[{Sanguinetti et~al.(2017)Sanguinetti, Kurani, and
  Davies}]{sanguinetti2017many}
Sanguinetti, A., Kurani, K., and Davies, J. (2017).
\newblock The many reasons your mileage may vary: Toward a unifying typology of
  eco-driving behaviors.
\newblock \emph{Transportation Research Part D: Transport and Environment}, 52,
  73--84.

\bibitem[{Teichert et~al.(2020)Teichert, Koch, and
  Ongel}]{teichert2020comparison}
Teichert, O., Koch, A., and Ongel, A. (2020).
\newblock Comparison of eco-driving strategies for different traffic-management
  measures.
\newblock In \emph{2020 IEEE 23rd International Conference on Intelligent
  Transportation Systems (ITSC)}, 1--7. IEEE.

\bibitem[{Wang and Nijmeijer(2015)}]{wang2015string}
Wang, C. and Nijmeijer, H. (2015).
\newblock String stable heterogeneous vehicle platoon using cooperative
  adaptive cruise control.
\newblock In \emph{2015 IEEE 18th International Conference on Intelligent
  Transportation Systems}, 1977--1982. IEEE.

\bibitem[{Wang et~al.(2019)Wang, Wu, and Barth}]{wang2019cooperative}
Wang, Z., Wu, G., and Barth, M.J. (2019).
\newblock Cooperative eco-driving at signalized intersections in a partially
  connected and automated vehicle environment.
\newblock \emph{IEEE Transactions on Intelligent Transportation Systems},
  21(5), 2029--2038.

\bibitem[{Wu et~al.(2015)Wu, He, Yu, Harmandayan, and Wang}]{wu2015energy}
Wu, X., He, X., Yu, G., Harmandayan, A., and Wang, Y. (2015).
\newblock Energy-optimal speed control for electric vehicles on signalized
  arterials.
\newblock \emph{IEEE Transactions on Intelligent Transportation Systems},
  16(5), 2786--2796.

\bibitem[{Yang et~al.(2016)Yang, Rakha, and Ala}]{yang2016eco}
Yang, H., Rakha, H., and Ala, M.V. (2016).
\newblock Eco-cooperative adaptive cruise control at signalized intersections
  considering queue effects.
\newblock \emph{IEEE Transactions on Intelligent Transportation Systems},
  18(6), 1575--1585.

\bibitem[{Zhai et~al.(2020)Zhai, Chen, Yan, Liu, and Li}]{zhai2020ecological}
Zhai, C., Chen, X., Yan, C., Liu, Y., and Li, H. (2020).
\newblock Ecological cooperative adaptive cruise control for a heterogeneous
  platoon of heavy-duty vehicles with time delays.
\newblock \emph{IEEE Access}, 8, 146208--146219.

\bibitem[{Zhao et~al.(2018)Zhao, Ngoduy, Shepherd, Liu, and
  Papageorgiou}]{zhao2018platoon}
Zhao, W., Ngoduy, D., Shepherd, S., Liu, R., and Papageorgiou, M. (2018).
\newblock A platoon based cooperative eco-driving model for mixed automated and
  human-driven vehicles at a signalised intersection.
\newblock \emph{Transportation Research Part C: Emerging Technologies}, 95,
  802--821.

\end{thebibliography}

\end{document}